\documentclass{article}
\usepackage{times} 
\include{psfig}
\usepackage{epsfig}

\begin{document}

\title{Morphological Complex Networks: Can Individual Morphology
Determine the General Connectivity and Dynamics of Networks?}

\author{Luciano da Fontoura Costa, \\
Instituto de F\'{i}sica de S\~ao Carlos. \\ 
Universidade de S\~ao Paulo, S\~{a}o Carlos, \\
SP, Caixa Postal 369, 13560-970, \\
Fone +55 16 3373 9858, FAX +55 1633 71 3616, \\
Brazil, luciano@if.sc.usp.br}

\date{18th March 2005}

\maketitle

{\bf ABSTRACT:} This article discusses how the individual
morphological properties of basic objects (e.g. neurons, molecules and
aggregates), jointly with their particular spatial distribution, can
determine the connectivity and dynamics of systems composed by those
objects.  This problem is characterized as a particular case of the
more general shape and function paradigm, which emphasizes the
interplay between shape and function in nature and evolution.  Five
key issues are addressed: (a) how to measure shapes; (b) how to obtain
stochastic models of classes of shapes; (c) how to simulate
morphologically realistic systems of multiple objects; (d) how to
characterize the connectivity and topology of such systems in terms of
complex network concepts and measurements; and (e) how the dynamics of
such systems can be ultimately affected, and even determined, by the
individual morphological features of the basic objects.  Although
emphasis is placed on neuromorphic systems, the presented concepts and
methods are useful also for several other multiple object systems,
such as protein-protein interaction, tissues, aggregates and polymers.

\begin{quotation}
'...the functional superiority of the human brain is \\
intimately linked up with the prodigious abundance \\
and unaccustomed wealth of forms of the so-called \\
neurons with short axons.' \\
\hspace{2cm} \emph{(Santiago Ram\'on y Cajal, Recollections of my life)}
\end{quotation}

\section{Introduction}

Look around and all you see is geometry.  Although string theory
suggests the universe is 10-dimensional, our perception is constrained
to 3 spatial dimensions which, jointly with time, accounts for the 4
traditional dimensions.  By contrast, a non-geographical graph (i.e. a
graph where the nodes do not have defined positions) lays in an
infinite dimensional space.  One of the most important consequences of
finite-dimensional spaces is the fact that the adjacencies imposed by
such spaces, jointly with the presence of spatial constraints
(e.g. impenetrability) and the fact that real objects are finite, tend
to severely limit long range connections and interactions.  For
instance, the only way to eat a pineapple is by cutting the fruit
somehow.  Even when interactions are allowed to go through space, as
is the case of electromagnetic fields and waves, such effects tend to
diminish steadily with distance.  At the same time as such constrains
limit our abilities, they are simultaneously responsible for
controlling the rate of dynamical changes in the universe, possibly
contributing to the enhancement of complexity.

One of the consequences of the spatial constraints and distance
metrics typically found in finite-dimensional spaces is the emergence
of shape and morphology (e.g.~\cite{Thompson:1917, Bookstein:1991,
Small:1996, Dryden_Mardia:1998, Pete:2000, CostaCesar:2001}).  For the
purposes of this work, a \emph{shape} will be henceforth understood as
any spatially limited and connected object~\cite{CostaCesar:2001},
while \emph{morphology} will stand for the study of shapes.  Systems
obtained by combining, through connection or not, objects or shapes
will be understood as \emph{morphic multiple object systems --- MMOS}.
Examples of shapes include the geographical space occupied by objects
such as molecules, cells, organs, and so on.  Combinations of
molecules give rise to multiple object systems such as materials,
combinations of cells to organs, while combinations of organs and
tissues produce living beings.  Observe that the definition of shape
is relative and somewhat arbitrary.  For instance, we may be
interested in the own shape of MMOS, such as the shape of the heart or
lungs, or in the shapes of each constituent object.

The \emph{shape-function} paradigm, whose origin is lost in the
history of science (going back to ancient Greece, especially
Pythagoras, at the very least), suggests that there is a close
interplay between shape and function in nature.  For instance, the
shape of proteins, together with field interactions, is essential for
defining their docking potential.  At the same time, the wing of an
airplane or a bird is essential for obtaining proper aerodynamics,
while the effectiveness of our lungs is directly related to the
intricacy of the bronchial channels, which allows enhanced surface
area of contact with air.  Although the shape-function paradigm can be
considered for the investigation of almost every natural phenomenon,
the current article concentrates attention on multiple object systems
which rely on the geometry of its basic components in order to obtain
proper behavior.  An immediate example of such systems is provided by
the skeletons of vertebrates, which are intrinsically suited to
perform the mechanical activities of such animals.  A more
sophisticate example, which will be central in our discussion, are the
biological neuronal networks such as those found in the mammals brain.

The brain is composed by a myriad of cells --- the neurons, whose
morphology is highly specialized to make \emph{selective} connections.
Interestingly, neurons do not connect indiscriminately between
themselves, but form an intricate system of short to long-range
connections with specific targets~\cite{Kandel:1995}.  In spite of
continuing efforts, the complete understanding on how neuronal
connections are established remains a considerable challenge.  Current
knowledge indicates that neuronal wiring is mediated by every sort of
field interactions, including electrical, neurotrophic factors,
gradients of ionic and molecular concentrations, gravity, among other
possibilities.  To any extent, the neuronal milieu represents a nice
example of how prolongations (i.e. dendrites and axons) are required
from neurons in order to overcome the adjacency constraints imposed by
the finite 3 spatial dimensions.  At the same time, the degree of
spatial occupancy and complexity of a dendritic tree is directly
related to its potential as a synaptic target for growing axons. The
immediate consequence of such effects is the fact that neuronal cells
tend to have a most diverse and specialized overall morphology,
ranging from simple bipolar cells to highly elaborated Purkinje
cells.

The complexity of neuronal shapes is immediately substantiated by the
fact that there is no current agreement between scientists regarding
the number of morphological classes of neurons, with competing
alternatives suggesting from 2 to thousands of types.  It is therefore
hardly surprising that the diversity of neuronal cells identified by
Ramon y Cajal in the beginning of the 20th century would be understood
by he, the father of modern neuroscience, as a possible explanation
for the complexity of human behavior and
intelligence~\cite{Cajal:1989}.  Since then, several evidences about
the relationship between neuronal shape and function have been
provided, including the congruence between the eletrophysiological
(i.e. functional) and morphological classes of retinal ganglion
cells~\cite{Boycott_Wassle:1974} as well as the correlation between
the structure of the receptive fields of those cells and the
respective morphology~\cite{Peichl_Wassle:1983}.  However, despite
such cumulative evidence about the interplay between neuronal shape
and function, most artificial neural networks, such as the
perceptron~\cite{Rosenblatt:1958} and Hopfield
models~\cite{Hopfield:1982}, have not incorporated the neuronal
morphology.

\begin{figure}
  \begin{center}
    \includegraphics[scale=.45]{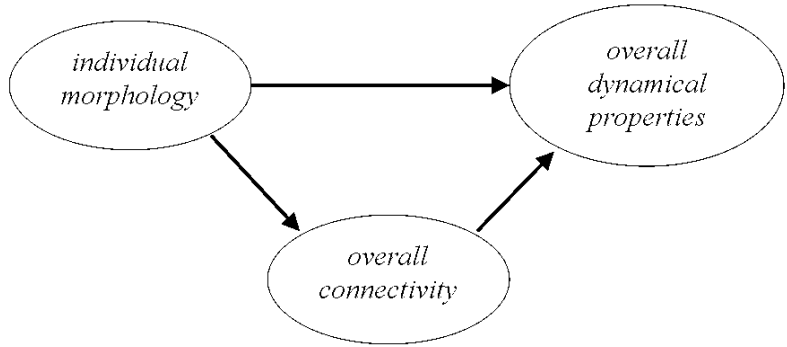}  
    \caption{The individual morphological features of basic
      objects can affect the overall dynamical properties of
      multiple object systems to which they belong either 
      directly or by determining the overall connectivity
      between the objects.  Very often, as in neuronal systems,
      the influences are two-sided, in the sense that the overall
      dynamics may also affect the connectivity and shape of
      the individual constituent objects.}~\label{fig:basic}
  \end{center}
\end{figure}

The mammals brain thus provides what is possibly the most compelling
illustration of a morphic multiple object system whose dynamics is, to
a substantial extent, the byproduct of the shapes of the individual
elements (e.g.~\cite{L_sibgrapi:1997}).  In other words, the shape of
the individual components becomes essential for obtaining patterns of
connectivity which are required for producing, through
electrophysiological means, function and behavior of great complexity.
Although \emph{complexity} remains a somewhat ellusive
concept~\cite{Gell_Mann:1995}, the primate brain is characterized by a
continuum of complexity flowing from the sophistication of the
\emph{shape} of its basic elements (i.e. the neuronal cells) to the
complexity of the emerging \emph{dynamics}, passing through the
complexity of \emph{connections} between such cells.  As illustrated
in Figure~\ref{fig:basic}, the geometrical properties of the
individual neuronal cells define the connectivity between those cells,
which ultimately produces complex dynamics.  Interestingly, such
effects are bilateral.  For instance, the shape of neuronal cells is
changed both as a consequence of stimuli presentation, which promotes
the appearance of new neuronal processes and synapses, as well as by
evolutionary forces acting during long periods of time.

Interesting as it can be, biological neuronal systems are but an
example of how the geometrical properties of individual components can
affect, or even determine, the emerging connectivity and dynamics of
multiple object systems.  This article provides an overview of such an
important and relatively overlooked paradigm from the following five
perspectives: (a) how to measure relevant geometrical properties of
shapes; (b) how to obtain stochastic models of classes of shapes; (c)
how to simulate systems of multiple morphological objects; (d) how to
characterize the connectivity and topology of such systems in terms of
complex network concepts and measurements; and (e) how the dynamics of
such systems can be ultimately influenced and determined by the
individual morphological features of the basic constituent elements.
Although emphasis is placed on neuromorphic systems, the problem of
protein interaction is also briefly discussed from the above stated
perspectives.

It should be noted that this work is relative to the oral presentation
of the author at the COSIN Final Meeting in Salou, Spain, which took
place in March 2005.  Emphasis is placed on integrating several
related previous developments by the author under the shape-function
paradigm, so that this text should by no means be treated as a
comprehensive survey to related topics.  As a consequence, great part
of the supplied references are to developments of the author over the
last decade, so that the reader is encouraged to refer to the supplied
references for a more comprehensive literature.

\section{Shape Measurement and Characterization}

Given any shape, it is possible to map its morphological properties
into a number of measurements, which can be suitably represented as a
\emph{feature vector}~\cite{CostaCesar:2001}, namely a vector in $R^N$
where each entry corresponds to each measurement of an object under
analysis.  For instance, given the shape of a neuronal cell, one may
be interested to quantify its volume,
orientations~\cite{L_scinst:1995}, fractal
dimension~\cite{L_velte:1999,LU101}, lacunarity~\cite{Gefen:1983,
Mandelbrot:1983, Allain_Cloitre:1991}, wavelet
features~\cite{L_robsproc:1997, L_biocyb:1998}, curvature and bending
energy~\cite{L_scientinstr:1997, L_velte:1999,LU101}, excluded volume
and autocorrelations~\cite{L_exclvol:2005}, as well as families of
Minkowski shape functionals~\cite{PRE_Marconi:2003}, to name but a few
among many other possibilities.

Given such a plurality of alternatives, a fundamental question arises:
what is the best set of measurements?  The answer to this question
should take into account the purpose of the measurements.  For
instance, in case we are interested in quantifying the overall cell
metabolism, a potentially useful measurement would be the total volume
of the cell.  Two of the most relevant situations in neuromorphology
concerns the choice of sets of measurements capable of: (a)
distinguishing between neuronal cell classes and (b) provide a
comprehensive representation of the cell geometry, in the sense that
the cell shape could be recovered from the measurements, i.e. the set
of measurements provides a \emph{complete} representation of the
neuronal shape.  However, even if the objectives are clearly stated in
each of these cases, the optimal set of features is still relative to
the specific types of neurons under analysis.  For example, in case
one wants to separate large from small cells, it may be enough to
consider the \emph{diameter} of the cells, namely the largest distance
between any pair of points obtained from the cell.  Yet, there are
some properties of measurements which are generally sought in both
situations (a) and (b) mentioned above, especially invariance to
translation and rotation.  Such properties, which are found in several
features such as area, perimeter, curvature, among others, are
important because often the shapes in MMOS can appear freely
translated and rotated and we do not want the measurement to take this
into account. Another relevant issue to be kept in mind is the fact
that the set of features should be minimal.

As the choice of features for shape representation will be covered in
more detail in the next section, we present in the following some
general reasonings on the selection of features for the purpose of
shape discrimination or geometrical pattern recognition (see, for
instance,~\cite{Duda_Hart:2001, CostaCesar:2001}).  Here the task
consists in, given one or more shapes, to organize then into classes
such that an object inside each class tends to be more similar to
other objects in that same class than to objects from the other
classes.  Observe that this general problem of pattern recognition has
some interesting similarities with the problem of community detection
in complex networks.  Note that we may or may not know the number of
classes, and that we may have or not have examples of objects from
each class.  In case we do have examples of objects from all the
classes, or information about the propeties of those objects, we say
we have a \emph{supervised} pattern recognition problem; otherwise, it
becomes a \emph{unsupervised} pattern recognition
case~\cite{Duda_Hart:2001, CostaCesar:2001}.  Needless to say, the
first type of problem is generally easier to be solved than the second
type.

Unfortunately, the choice of features for pattern recognition is still
largely subjective in the sense that there are no definite rules which
can be applied for that end.  The problem is complicated by the fact
that there is an infinite number of features allied to the fact that a
large number of measurements do not necessarily lead to better
discrimination (e.g.~\cite{CostaCesar:2001}).  In practice, one has to
rely on previous experience about the problem under analysis, as well
as with the properties of the several available features.  The
remainder of this section presents some brief remarks about some
representative shape measurements and their specific interpretation
and characteristics.

{\bf Hierarchical angles and lengths:} Provided the shape can be
organized along hierarchical levels, namely as a branching pattern, an
interesting set of measurements can be obtained by taking into account
the angles and arclenghts of the constituent segments
(e.g.~\cite{L_regina:1997, L_regina:2002}).  The hierarchical
decomposition, which is more viable when applied to branching
patterns, can be obtained by applying skeletonization approaches as
discussed in~\cite{L_santorini:1999, L_eletters:1999,
L_andetal:2000,L_falcao:2002,L_skel:2000, Costa_Muti:2003,
L_jneu:2005} or by curvature-based methods as described
in~\cite{L_dendro:1999}.  In the case of neuronal cells or shapes
which have a body, it is important to detect such a
body~\cite{L_jneu:2005} in order to have the correct identification of
the beginning of each process (i.e. a dendrite or axon).

{\bf Curvature:} Quantifying the rate of change of the tangent angle
of a given curve, the curvature provides invariance to translation and
rotation and is complete (i.e. inversible) up to such transformations.
The identification of high curvature points provide valuable resources
for identifying extremities along shapes~\cite{L_scinst:1995,
L_dendro:1999}.  An effective approach to numerical curvature
estimation has been described~\cite{L_robcurv:1996,
L_scientinstr:1997, L_dendro:1999, L_velte:1999,
CostaCesar:2001,L_dsp:2003} which estimates the shape derivatives
required for curvature calculation by using the derivative property of
the Fourier transform, combined with Gaussian regularization
(i.e. low-pass filtering).  The histogram of curvatures of a shape, as
well as its respective statistical moments, may also provide valuable
information for shape characterization and discrimination.

{\bf Bending energy:} Given a differentiable two-dimensional shape,
its bending energy can be shown to be proportional to the sum of the
squared curvature values.  Therefore, the more intrincated the shape
is, the higher its bending energy will result.  The bending energy has
been used with encouraging success for classification of neuronal
cells~\cite{L_scientinstr:1997}.

{\bf Fractal dimension:} The several available definitions of fractal
dimension provide an interesting alternative for quantifying the
`complexity', spatial distribution or spatial coverage of a given
shape (see, for instance,~\cite{CostaCesar:2001, LU101}).

{\bf Lacunarity:} Introduced in order to complement the fractal
dimension~\cite{Gefen:1983, Mandelbrot:1983, Allain_Cloitre:1991}, the
lacunarity quantifies the degree of translational invariance of a
pattern.  Typically, the higher the lacunarity, the less translational
invariant the shape is.  The traditional way to calculate the
lacunarity involves sliding a window along the whole image while
estimating the mass comprised by the window.  Several sizes of windows
are usually considered, yielding the lacunarity as function of the
windows size.  A recent study~\cite{L_self:2005} has shown that
translational and rotational invariance can be achieved by using
circular windows centered only at the object points, not through the
whole workspace.

{\bf Minkowski functionals:} As implied in the name, shape functionals
are maps from a given shape to scalar values.  Minkowski functionals
are a special class of functionals which are additive, motion
invariant and continuous~\cite{Michiel1:2000, Michiel2:2001,
Mecke:1998}. In the plane, such functionals comprise the perimeter,
area and Euler number, corresponding to the number of holes in the
object under analysis.  Minkowski functionals have been successfully
applied to the characterization of neuronal
shape~\cite{PRE_Marconi:2003, L_morphalt:2004}.

{\bf Critical percolation densities:} Given a shape and an empty
workspace, the progressive superposition of this shape at uniformly
distributed random positions along the workspace eventually lead to
percolation, which can be identified by looking for a cluster of
shapes extending from side to side of the
workspace~\cite{Percolation:2003}.  The average density of shapes
observed when percolation occurs, called henceforth the critical
percolation density, has been suggested~\cite{Percolation:2003} as the
most direct measurement of the potential of the given shape for
establishing connections, leading to promising discriminative
potential.  Interesting previous investigations of percolation in
neuronal systems, where the neuronal shape was simplified as circles,
have been reported in~\cite{Ooyen:1995}.  The critical percolation
density can also be estimated for ballistic deposition, as
investigated in~\cite{L_ballistic:2005}.  In the case the shapes
change with time, as is the case with growing neurons, it is possible
to define a critical percolation density which is a direct consequence
of the growth dynamics and spatial distribution of the
cells~\cite{L_regin:2004}.  In the case of a collection of not
necessarily connected and static objects, it is still possible to
force percolation through some imposed growth dynamics, such as
parallel dilation~\cite{L_bioinfo:2005, L_active:2004, L_active:2005}.

{\bf Relative shape measurements:} While all the above measurements
are specific only to the shape of interest, it is also important to
consider geometrical features that also take into account
environmental constraints such as the presence of other objects, as
well as fields and other types of interactions.  One example of such
measurements would be to consider the angle difference between the
tangent field orientation along the object, which is a measurement
intrinsically related to the shape, and the electric field orientation
at those same points induced by the presence of surrounding objects.
Low averages of such ratios would therefore indicate that the shape in
question is highly affected by the external objects.  Such
measurements are particularly important for studying and modeling MMOS
whose basic objects are known to be affected by environmental
influences, as is the case with neuronal systems.

\section{Stochastic Shape Modeling and Synthesis}

In this section we discuss the problem of how to produce, by
mathematic-computational means, artificial instances of shapes of a
given category.  For instance, one may be interested in obtaining a
collection of neuronal cells which are statistically equivalent to
those of a certain biological class (e.g. alpha cat retinal gangion
cells).  By \emph{statistical equivalence} we mean that the
synthetized cells will produce the same statistical distribution of
values of geometrical measurements.  Observe that, given the almost
unlimited statistical variability of the properties of natural shapes,
even within the same category of shapes, the problem of shape
equivalence in the sense of exactly reproducing a given shape is of
minor interest.  On the contrary, it becomes more important to devise
synthesis methodologies capable of incorporating the stochastic
variability found in nature.

Two important situations arise: (i) the original shapes evolve without
influences from the environment and (ii) the shapes are affected by
the environment.  The problem of stochastic shape synthesis in these
two situations are discussed in the following subsections.

\subsection{Absence of environmental influences}

In this case, the shapes of interest are the result of a growing
dynamics which takes into account only internal constraints imposed by
the physics/biology of the object.  While there are no natural
examples of such growth dynamics, this approximation provides a first,
simple approach to modeling and simulating shapes.  In the following
we illustrate this kind of strategy with respect to the simulation of
neuronal outgrowth~\cite{L_regina:1997, L_regina:2002}.  Other
approaches to neuromorphic modeling include those adopting differential
equations to model cytoskeleton outgrowth ~\cite{Hentsch1:1994,
Hentsch2:1998, L_dania:2003, L_andneuro:2001}, methods based on
Hillman's~\cite{Hillman:1979} set of quantitative anatomical
correlations~\cite{GA_hill1:2000, GA_hill2:2001}, activity and
competition during process outgrowth~\cite{Ooyen:1995, Ooyen:1999,
Ooyen:2001, Ooyen:2004} as well as hidden Markov
models~\cite{GA_hipp:2005}.

\begin{figure*}
  \begin{center}
    \includegraphics[scale=.4]{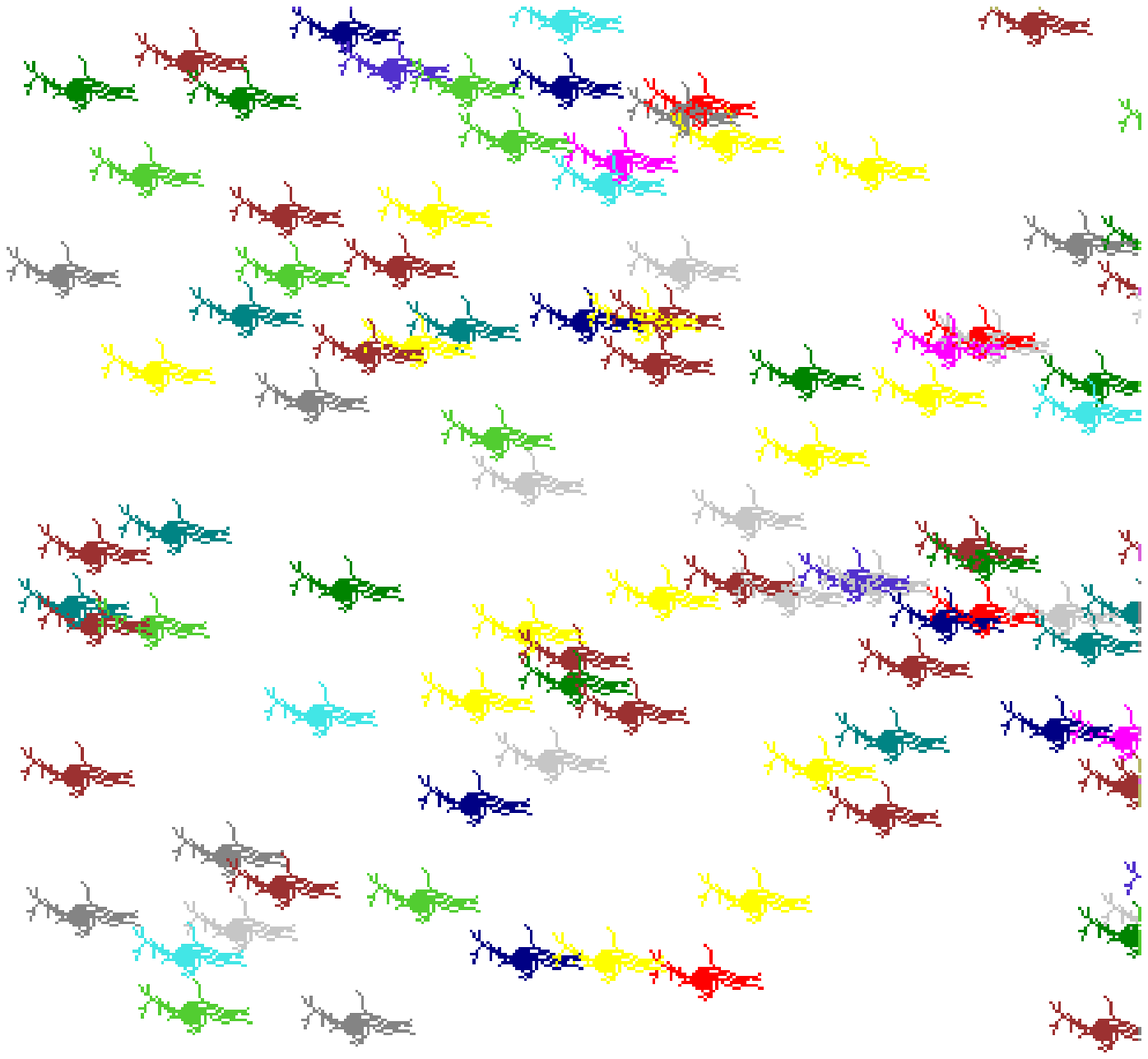}   \hspace{0.5cm}
    \includegraphics[scale=.4]{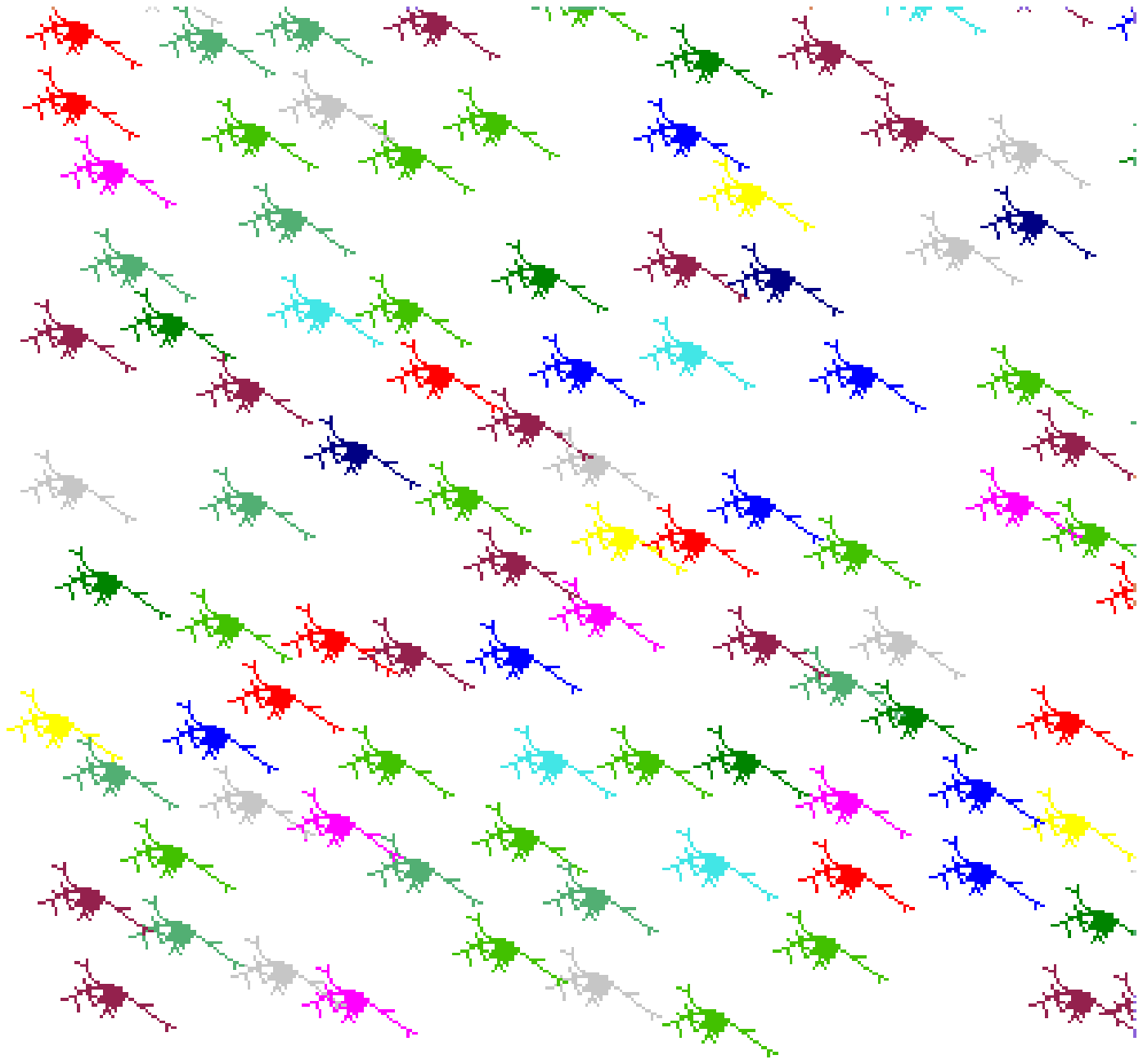}  
    (a)   \hspace{5.5cm}     (c)   \vspace{0.5cm} \\
    \includegraphics[scale=.4]{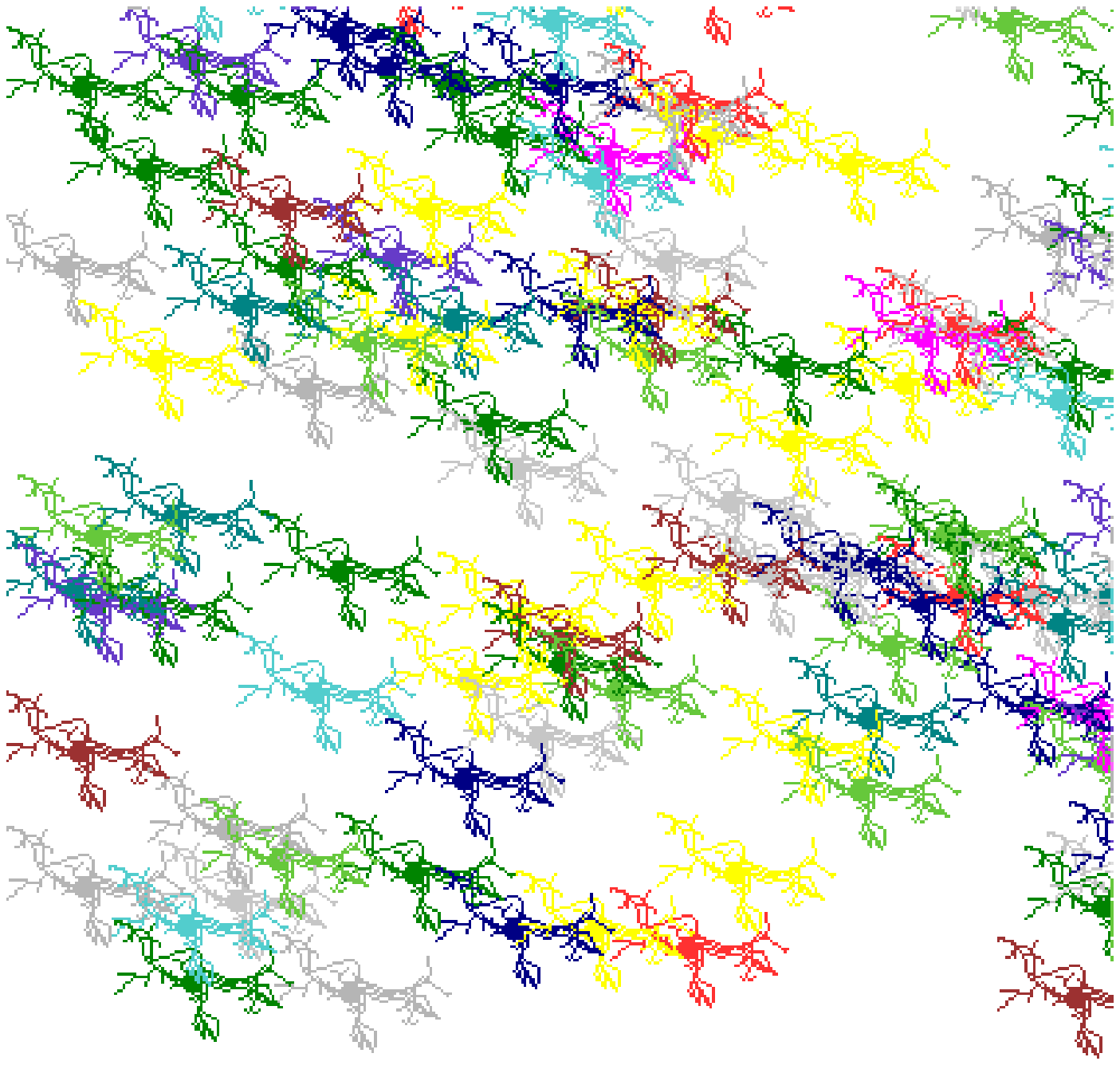}   \hspace{0.5cm}
    \includegraphics[scale=.4]{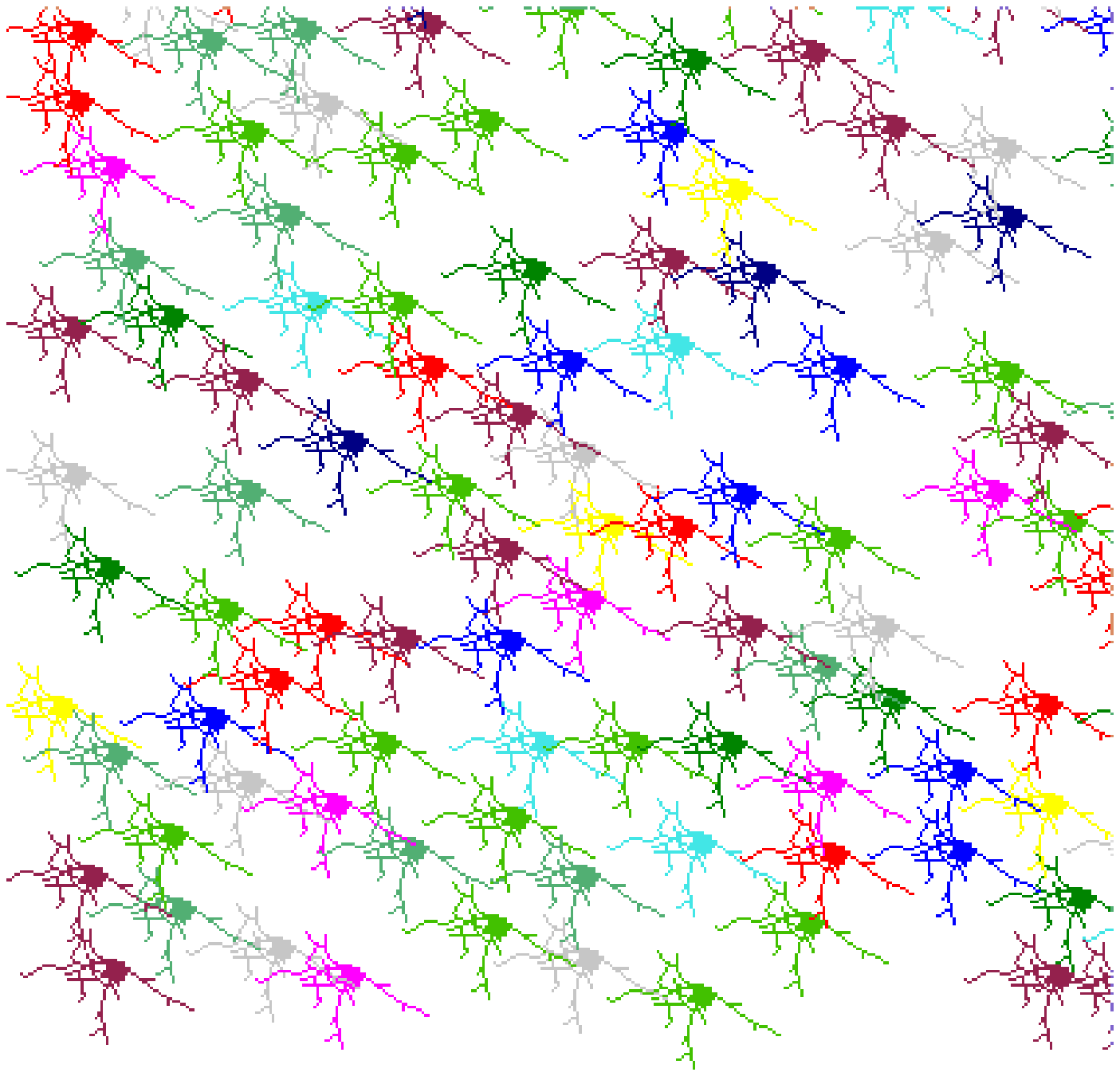}  
    (b)   \hspace{5.5cm}     (d) 
    \caption{Three stages along neuronal growth simulations for alpha
    (a-b) and beta (c-d) cells.  In this example, all cells are
    identical.}~\label{fig:growth}
  \end{center}
\end{figure*}

The overall idea of the conditional probability
model~\cite{L_regina:1997, L_regina:2002} involves obtaining
representative measurements from the real biological cells, such as
the angles and lengths along the branching hierarchies, represented as
conditional densities, and then Monte Carlo sampling such
distributions in order to produce synthetic neuronal cells.
Figure~\ref{fig:growth} illustrates three stages along the development
of simulated alpha and beta cells.  In both cases, percolation is
observed between the growth stages illustrated in the figure. Although
this example considers identical cells, which reduces the variance of
the percolation critical density, simulations assuming different
neuronal cells have are also important and have been investigated.

Given a collection of adult neuronal cells, each of them can be imaged
in 2D or 3D by using several types of microscopy, including optical
transmission and confocal.  Although the following discussion is
limited to 2D neuronal shapes (several neuronal types, such as retinal
ganglion cells, are mostly planar), the extension to 3D is immediate.
The image of the neuronal cell can then be processed in order to
obtain the points belonging to the cell and the points belonging to
the background.  The contour of such cells is particularly important
and can be obtained by using edge detection and contour following
algorithms (e.g.~\cite{CostaCesar:2001}).  The particularly important
points of the neuronal ramification, namely the extremity and
branching points, can be obtained by applying a curvature-based
procedure~\cite{L_linseg:1995, L_dendro:1999} or skeletonization
algorithms~\cite{L_andetal:2000, CostaCesar:2001}.  The angles and
arclengths along each segment along the hierarchies of the
ramifications can then be obtained by using simple analytical geometry
concepts and methods, so that conditional probability densities of the
angles and lengths can be obtained.  As discussed in the previous
section, such measurements provide, by themselves, a naturally
suitable characterization of the neuronal geometry.  Note that the
number of hierarchies considered in the conditional densities is
limited by the amount of neuronal samples, in the sense that the
consideration of several hierarchies implies higher dimensionality of
joint densities which require larger numbers of cells in order to
obtain statistically representative results (i.e. a statistical
sampling problem). Finally, the conditional densities can be Monte
Carlo sampled in order to produce a virtually infinite number of
synthetic cells whose geometrical features will lead to statistical
densities equivalent to those which were sampled. 

\subsection{Presence of environmental influences}

The more realistic situation of shape development under environmental
influcences requires a modeling approach capable of integrating
\emph{both} internal and external influences, because the former are
always present in physical and biological systems.  Often, external
influences manifest themselves in terms of \emph{fields} or
\emph{waves}.  For instance, the growth cones (namely the tips of
growing axons) are known to follow gradient fields derived from
neurotrophic growth factors, ionic concentrations, electric, and even
gravitational fields.  Therefore, it is reasonable to incorporate the
effect of such fields in neuronal outgrowth by adding to the instant
growth velocity (implied by internal influences) a vector component
parallel to the external field (e.g.~\cite{L_consul:1998,
L_andneuro:2001, L_dania:2003}).  It has been
verified~\cite{GA_repuls:2003} that the incorporation of a vector
component pointing away from the cell soma also accounts for
biological realism in neuronal outgrowth.

Because the neuronal activity during neuronal development can also
have effects in shaping the neuronal morphology, it is interesting to
consider such dynamics in simulations.  We can use the recently
introduced Sznajd complex networks~\cite{L_sznaj:2005}, whose topology
is determined by correlations between the neuronal
connectivity/activity, in order to implement Hebbian dynamics, where
the most active connections are reinforced.  Another effect to be
considered is the fact that waves of ionic concentration have been
verified to induce neuronal spikes~\cite{L_rpozn2:2001}.  Therefore,
it would be interesting to include such waves in neuronal outgrowth
simulations.

\section{Synthesis of Morphic Multiple Objects Systems and their 
Representation and Characterization in Terms of Complex Networks}

Provided we have implemented the means to grow individual shapes, they
can be combined in order to produce morphic multiple object systems.
The most natural means to do that involves starting the growth at $N$
seeds distributed spatially according to some desired statistical
density.  In the case of neurons, it is often reasonable to assume a
uniform distribution of cells (see~\cite{Costa_APL1:2005} for a
discussion on the measurement of spatial dispersion in neuronal
mosaics), which are then grown by using a combination of the
techniques discussed in the previous section.  Every time an axon gets
with a pre-specified distance from a dendrite, it may be connected to
that dendrite with some probability which may be a function of the
distance as well as other factores (e.g. diameter of the axon and type
of cells, among many other possibilities)~\cite{L_regina:1997,
L_regina:2002}.

Fortunately, MMOSs can be effectively represented in terms of
graphs/networks and their respective adjacency or weight
matrices~\cite{L_regina:1997, Costa_BM:2003, L_analyt:2003,
L_regin:2004}.  As such systems grow, it is possible to keep track of
every new connection (a synapse in the case of neuronal systems),
which can be updated into the adjacency or weight matrix of the graph
used to represent the MMOS.  In the case of neuronal networks, such a
representation is almost complete, except for the fact that it does
not take into account fields and waves emmanating from the
environment~\cite{L_rpozn2:2001}.  However, such effects can be easily
incorporated by using geographical networks for representing the
system, therefore providing information about the spatial position of
the nodes and allowing the determination of the fields at those
points.

One of the most interesting perspective in applying complex network
measurements for characterization and analysis of growing MMOS regards
obtaining each measurement as a function of time or development
epochs.  For instance, following traditional complex network
measurements such as the node degree and clustering coefficient in
terms of time can provide valuable information about the local
connectivity of the objects as the system evolves.  Virtually every
complex network measurement will have some additional aspect to say
about the evolving connectivity.  More global information about the
system connectivity can be gathered by using the recently introduced
hierarchical versions of the node degree and clustering coefficient,
as well as several new hierarchical
measurements~\cite{Costa_Hier:2004, Costa_Hier:2005, Costa_Gener:2004}
and wiring lengths between neuronal modules~\cite{L_topo:2005}.  The
application of community finding algorithms can also provide
particularly interesting information with implication to the
functional dynamics of the MMOS under analysis, with important
implications to the evolution of synchronization and attractors
formation.  Another promising perspective is the quantification of the
cycles of several lengths in the growing networks, which will also
have important implications for the system dynamics.

The most direct quantification of the evolution of the overall
connectivity of the system is provided by the critical percolation
density.  We have applied static and dynamics, natural and forced
percolations~\cite{L_active:2004} in order to investigate the
influence of the individual shape of neurons to the overall
connectivity~\cite{Percolation:2003, L_regin:2004}.  In investigations
considering alpha and beta cat ganglion cells with normalized sizes,
we observed that more complex cells tend to percolate sooner.
However, the critical percolation density or time (in the case of
growing systems with constant density) has been verified to be a
function of several other parameters, such as the neuronal cell
elongation, the excluded volume, as well as the straightness of the
neuronal processes.  Further investigations are required in order to
establish a clear relationship between the individual object
properties, as well as their growth dynamics, with the final overall
MMOS connectivity.

\section{Consequences on the Overall System Dynamics}

Because the individual morphological properties of the objects have
been found to determine, jointly with their spatial distribution, both
the local and global connectivity of MMOSs, such geometrical features
will also be inexorably related to the emmerging dynamics of the
system.  After investigating the effect of using scale free
connectivity in Hopfield networks~\cite{Stauffer_Costa:2003,
Costa_Stauffer:2003}, we have been investigating how the individual
neuronal shape can affect neuronal dynamics~\cite{Costa_BM:2003}.  We
have quantified the memory potential of the neuronal systems in terms
of the \emph{overlap}, which is related to the number of correct
recovered bits, between the originally trained pattern and the pattern
recovered (starting from a perturbated version of the original
pattern) after $P$ patterns had been trained into the neuronal system.
Considering small neurons with the same area, we have
found~\cite{L_analyt:2003} that such a memory potential is highly
dependent on the individual neuronal shape, with shapes exhibiting
ramified and broader distribution of mass (i.e. neuron-like) tending
to perform substantially better than simple shapes such as bars and
crosses.  The distribution of complex eigenvalue of the adjacency
matrices obtained with neuron-like cells have resulted less
degenerated, which is known to be associated with better recall
capabilities~\cite{Haykin:1999}.

\section{Conclusions and Future Works}

The present text provided an integrated review and discussion on the
problem of relating shape and function through connectivity in morphic
multiple object systems (MMOS).  The main paradigm here is the fact
that the geometric features at the individual level can play a
decisive role in defining the overall connectivity of the system, and
consequently its dynamical operation.  While the quantification of
geometrical features of the individual objects can be achieved by
considering a variety of measurements, we have argued that the
resulting connectivity of the respective MMOS can be effectively
quantified and modeled in terms of complex network concepts and
measurements.

Although the shape/function paradigm has been clearly identified and
investigated in preliminary works, a series of important problems
remain open to research.  Of particular interest would be to identify
how each of the main shape measurements are correlated to the overall
connectivity, as quantified by complex network measurements.  It would
also be interesting to develop further analytical models such as that
presented in~\cite{L_analyt:2003} so that the several topological
properties of the MMOS can be foreseen from the geometrical features
of the constituent objects.  One particularly interesting approach
would be to try to relate the fractal dimension of the individual
neuronal cells with the fractal dimension of the overall topology of
complex networks~\cite{Song_etal:2005}, and then with scale-free laws
and other properties of the emmerging connectivity and activity
dynamics of the respective networks.  Other promising perspectives
include the analysis of recall in neuronal systems other than
Hopfield, as well as the quantification of synchronization between
cells as the MMOS grows.  The most realistic situation where the
connectivity dynamics is related to the network activity dynamics
along time and space remains one of the most interesting and
challenging situation for simulations and investigations.  We are
working on such issues and would highly appreciate to consider
collaborations.

\vspace{0.5cm} 
{\bf Acknowledgments:} Luciano da F. Costa is grateful to
Regina C. Coelho (UNIMEP, Brazil), who kindly supplied the simulation
data shown in Figure 2, and to HFSP RGP39/2002, FAPESP
(proc. 99/12765-2) and CNPq (proc. 301422/92-3) for financial support.

\bibliographystyle{unsrt}
\bibliography{salou}

\end{document}